\documentclass[letterpaper, 10 pt, conference]{ieeeconf}  

\renewcommand{\baselinestretch}{0.99}
\usepackage[utf8]{inputenc}
\usepackage[T1]{fontenc}
\usepackage{breqn}
\usepackage{amsmath}
\usepackage{amsfonts}
\usepackage{amssymb} 
\let\labelindent\relax
\usepackage{enumitem}
\usepackage[normalem]{ulem}

\newtheorem{assumption}{Assumption}
\usepackage{algorithm} 
\usepackage{algpseudocode} 
\usepackage{pgfplots} 
\pgfplotsset{compat=newest} 
\pgfplotsset{plot coordinates/math parser=false} 
\newlength\figureheight 
\newlength\figurewidth 
\pgfplotsset{
    every axis plot post/.style={
        line join=round
    }
}

\usepackage{tikz}
\usepackage{tikzscale}

\definecolor{myorange}{cmyk}{0,0.35,0.85,0} 
\definecolor{mypurple}{cmyk}{0.5,1,0,0} 

\definecolor{matblue1}{rgb}{0,0.4470,0.7410}
\definecolor{matred1}{rgb}{0.85,0.325,0.098}
\definecolor{matyel1}{rgb}{0.9290, 0.6940, 0.1250}
\definecolor{matpur1}{rgb}{0.4940, 0.1840, 0.5560}
\definecolor{matgre1}{rgb}{0.4660, 0.6740, 0.1880}
\definecolor{matblue2}{rgb}{0.3010, 0.7450, 0.9330}
\definecolor{matred2}{rgb}{0.6350, 0.0780, 0.1840}
\definecolor{matgrey1}{rgb}{0.5, 0.6, 0.7}
\definecolor{matpink1}{rgb}{1, 0.07, 0.65}
\definecolor{matblue3}{rgb}{0.07, 0.62, 1}
\definecolor{gray09}{rgb}{0.9, 0.9, 0.9}

    \definecolor{mblue}{rgb}{0,0.447,0.741}
    \definecolor{mred}{rgb}{0.85,0.325,0.098}
    \definecolor{myellow}{rgb}{0.9290,0.6940,0.1250}
    \definecolor{mmagenta}{rgb}{1,0,1}
    \definecolor{mgreen}{rgb}{0.4460,0.6740,0.1880}
    \definecolor{mgrey}{rgb}{0.6,0.6,0.6}
    \definecolor{mpurple}{rgb}{0.4940, 0.1840, 0.5560}
    
    \tikzset{cross/.style={cross out, draw=black, minimum size=2*(#1-\pgflinewidth), inner sep=0pt, outer sep=0pt}, cross/.default={1pt}}
    
    \usetikzlibrary{shapes}

\newcommand{\blackdash}{\raisebox{2pt}{\tikz{\draw[-,black,dashed,line width = 0.9pt](0,0) -- (3mm,0);}}}

\newcommand{\blueline}{\raisebox{2pt}{\tikz{\draw[-,matblue1,solid,line width = 0.9pt](0,0) -- (3mm,0);}}}
\newcommand{\redline}{\raisebox{2pt}{\tikz{\draw[-,matred1,solid,line width = 0.9pt](0,0) -- (3mm,0);}}}

\newcommand{\yelline}{\raisebox{2pt}{\tikz{\draw[-,matyel1,solid,line width = 0.9pt](0,0) -- (3mm,0);}}}

\newcommand{\greyarea}{\raisebox{0pt}{\tikz{\draw[-,gray09,solid,line width = 4pt](0,0) -- (3mm,0);}}}

\usetikzlibrary{external} 
\usetikzlibrary{spy}
\usepackage{graphicx}      
\IEEEoverridecommandlockouts                              

\title{\LARGE \bf
Nonlinear Bayesian Identification for Motor Commutation:\\Applied to Switched Reluctance Motors}

\author{Max van Meer$^{1}$, Rodrigo A. González$^1$, Gert Witvoet$^{1,2}$, Tom Oomen$^{1,3}$
\thanks{$^{1}$Max van Meer (e-mail: m.v.meer@tue.nl), Rodrigo A. González, Gert Witvoet, and Tom Oomen are with the Control Systems Technology section, Department of Mechanical Engineering, Eindhoven University of Technology, The Netherlands. This work is part of the research program VIDI with project number 15698, which is (partly) financed by the Netherlands Organisation for Scientific Research (NWO). In addition, this research has received funding from the ECSEL Joint Undertaking under grant agreement 101007311 (IMOCO4.E). The Joint Undertaking receives support from the European Union's Horizon 2020 research and innovation program.}
\thanks{$^{2}$Gert Witvoet is also with the Department of Optomechatronics, TNO, Delft, The Netherlands.}
\thanks{$^{3}$Tom Oomen is also with the Delft Center for Systems and Control, Delft University of Technology, Delft, The Netherlands.}
}

\usepackage{afterpage}

\usepackage{etoolbox}

\newlength{\algorithmspace} 
\BeforeBeginEnvironment{algorithm}{%
  \setlength{\textfloatsep}{\algorithmspace} 
  \setlength{\intextsep}{\algorithmspace}    
  \setlength{\abovecaptionskip}{\algorithmspace} 
  \setlength{\belowcaptionskip}{\algorithmspace} 
}

\AfterEndEnvironment{algorithm}{%
   \setlength{\abovecaptionskip}{-2pt}
   \setlength{\belowcaptionskip}{-3pt}
}

  \usepackage{eso-pic}
  \AddToShipoutPictureBG*{%
  \AtPageUpperLeft{%
  \setlength\unitlength{1in}%
  \hspace*{\dimexpr0.5\paperwidth\relax}
  \makebox(0,-1.00)[c]{
  \begin{tabular}{c c}
  Max van Meer, 
  Nonlinear Bayesian Identification for Motor Commutation: Applied to Switched Reluctance Motors \\
  Accepted for 
  {\em 62nd IEEE Conference on Decision and Control},
  Singapore, 2023,
  uploaded to arXiv \today \\
  \end{tabular}}}}

\begin{document}

\maketitle
\thispagestyle{empty}
\pagestyle{empty}

\begin{abstract}
Switched Reluctance Motors (SRMs) enable power-efficient actuation with mechanically simple designs. This paper aims to identify the nonlinear relationship between torque, rotor angle, and currents, to design commutation functions that minimize torque ripple in SRMs. This is achieved by conducting specific closed-loop experiments using purposely imperfect commutation functions and identifying the nonlinear dynamics via Bayesian estimation. A simulation example shows that the presented method is robust to position-dependent disturbances, and experiments suggest that the identification method enables the design of commutation functions that significantly increase performance. The developed approach enables accurate identification of the torque-current-angle relationship in SRMs, without the need for torque sensors, an accurate linear model, or an accurate model of position-dependent disturbances, making it easy to implement in production. 
\end{abstract}

\section{INTRODUCTION}
\begin{figure}[t]
    \centering
    \vspace{3pt}
    \includegraphics[width=0.5\linewidth]{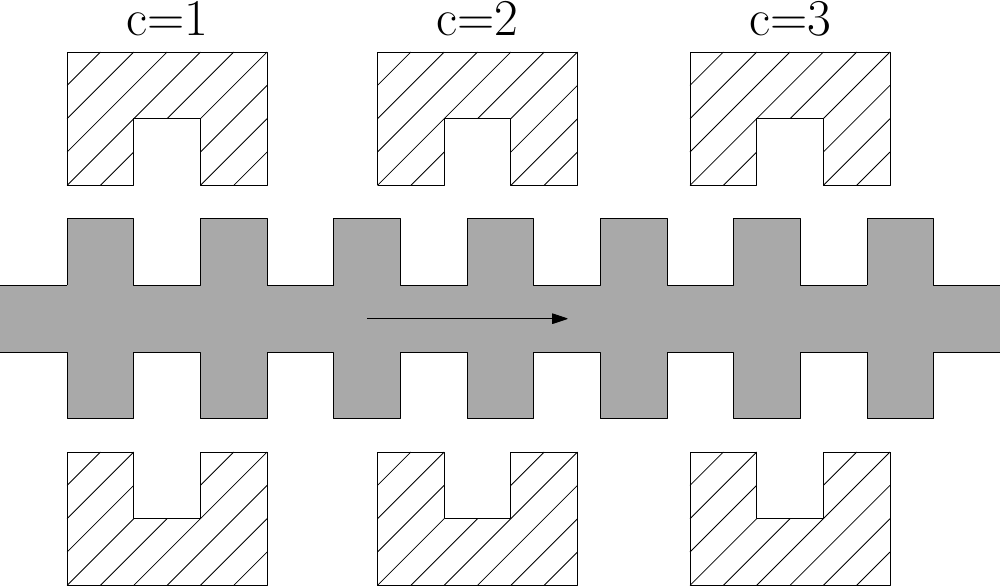}
    \caption{Working principle of an SRM with three coils. By applying a current to one of the three coils, a magnetic field is created, attracting the closest rotor tooth. The direction and magnitude of the resulting torque on the rotor depend on the applied current and the rotor position.}
    \label{fig:srm}
\end{figure}

Switched Reluctance Motors (SRMs) are electric actuators that have gained increased attention in the past decade due to their power efficiency, mechanical simplicity, and lack of permanent magnets \cite{Katalenic2013}. Since the power is applied to the stator instead of the rotor, no commutator is required, simplifying the mechanical design but complicating control, as current needs division among the coils, see Figure \ref{fig:srm}.

\begin{figure}
    \centering
    \includegraphics[width=\linewidth]{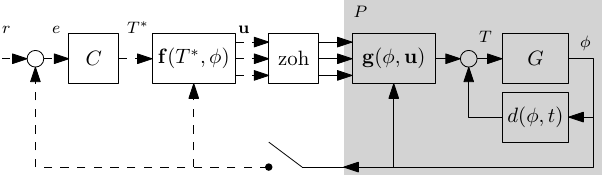}
    \caption{Control scheme for an SRM $P$. The nonlinear system $P$ is linearized by designing an $\mathbf{f}$ satisfying $\mathbf{g}\mathbf{f}\approx1$, which enables the use of a linear feedback controller $C(s)$. Recall that $d$ is assumed small compared to $\mathbf{g}\mathbf{f}$.}
    \label{fig:controlscheme}
\end{figure}

Linear feedback control is a powerful tool in a wide range of applications since it enables a convenient assessment of stability and performance \cite{Skogestad2005}. Hence, it is desirable to design a so-called commutation function \cite{Wang2016} that inverts the nonlinear torque-current-angle relationship by computing coil currents for a certain desired torque and rotor position, effectively linearizing the system and enabling linear feedback control, see Figure \ref{fig:controlscheme}. For a given SRM, infinitely many possible commutation functions exist that invert the nonlinear torque-current-angle relationship, and this design freedom can be exploited to enforce a trade-off on desired properties such as peak currents or power consumption through heuristic \cite{Vujicic2012} or optimization-based methods \cite{Meer2022}. 

Imperfect commutation functions result in torque ripple, a position-dependent mismatch between desired and true torque. Several methods exist to mitigate torque ripple, e.g., using linear feedback \cite{Kramer2020} or using spatial repetitive control \cite[Chapter 5]{Mooren2022Thesis}. Such methods typically sacrifice the design freedom in commutation design by merely requesting more torque from an imperfect commutation function. If instead, an underlying model of the torque-current-angle relationship is available, then torque ripple can be reduced while also enforcing desired properties on the current waveforms.

The identification of the torque-current-angle relationship in SRMs is highly complicated in the absence of torque sensors. Indeed, many applications that favor the use of SRMs because of their cost-effectiveness may not equip additional sensors that increase cost and complexity, e.g., in satellite communication terminals \cite{Kramer2020}.

Numerous nonlinear system identification techniques \cite{Schoukens2019a} exist to identify nonlinear systems. These approaches can be distinguished into two categories: $(i)$ methods that yield physical models and $(ii)$ methods that yield black-box models. Identification methods yielding physical models rely on a priory knowledge of the nonlinear structure of the system to identify system parameters. Black-box approaches do not require knowledge of the nonlinear structure to obtain an accurate model, at the cost of high model order and low interpretability.
The highly complex structure of nonlinear SRM dynamics, influenced by position-dependent disturbances like friction and magnetic saturation, makes specifying a suitable model structure exceedingly difficult. At the same time, an interpretable model of the SRM dynamics is desired, since a commutation function only inverts the nonlinear torque-current-angle relationship. When the full SRM model is a black box, the commutation function design is complicated, if not impossible. Indeed, when a black box model simulates only the relation between input (currents) and output (rotor angle), the unmeasured generated torque is a hidden state that is not accessible for commutation function design. 

Although existing nonlinear system identification methods are effective in various applications, they are limited in identifying SRMs. Physical model identification requires intricate model structure knowledge, and black-box methods complicate commutation function design. Therefore, this paper aims to obtain a physical model of only the nonlinear torque-current-angle relationship of SRMs in the absence of torque sensors, while being robust to unmodeled nonlinear disturbances. This is achieved through specific choices in the experiment design, that enable the posing of a Bayesian estimation problem. The contributions of this paper are: \begin{enumerate}[label={C\arabic*:}]
    \item An identification method for SRMs in the absence of torque sensors is developed. By performing several closed-loop experiments with commutation functions that are imperfect by design, a physical model of the torque-current-angle relationship is obtained. 
    \item A simulation example demonstrates that the method is robust to unmodeled position-dependent disturbances. 
    \item Experimental results show that the identification method leads to superior tracking performance in a case study.
\end{enumerate}

The paper is organized as follows: Section \ref{sec:problem} describes the problem; Sections \ref{sec:method} and \ref{sec:implementation} detail the method and implementation; Sections \ref{sec:simulation} and \ref{sec:experiments} present simulation and experimental results; and Section \ref{sec:conclusions} concludes.
\section{PROBLEM DESCRIPTION}\label{sec:problem}
In this section, the considered problem is defined. First, the actuation principle of Switched Reluctance Motors is explained. Next, the control of SRMs is described, and finally, the problem formulation of identifying the torque-current-angle relationship is presented.
\subsection{Switched Reluctance Motor dynamics}
Switched Reluctance Motors, as illustrated schematically in Figure \ref{fig:srm}, exhibit a nonlinear relationship between torque, current, and rotor angle. An SRM with $n_t$ teeth and $n_c$ coils, without magnetic saturation, is modeled as
\begin{equation}
T_c(\phi,i_c) = \frac{1}{2} \frac{\textnormal{d} L_c(\phi)}{\textnormal{d} \phi} i_c^2,
\end{equation}
where $T_c$ is a torque applied to the rotor by magnetizing coil $c\in \{1,\ldots,n_c\}$ with a current $i_c$, and $L_c(\phi)$ is the phase inductance, which varies periodically with the rotor position $\phi$, with spatial period $\frac{2\pi}{n_t}$. Hence, the total torque applied to the rotor at time $t$ is given by \begin{align}\label{eq:nonlin_plant}
T(t) &= \mathbf{g}(\phi(t)) \mathbf{u}(t) + d(\phi,t),\\
\mathbf{g}(\phi) & :=\frac{1}{2}\frac{\textnormal{d}}{\textnormal{d}\phi}[L_1(\phi),\ldots,L_{n_c}(\phi)], \\
\mathbf{u}{(t)} & := [i_1(t)^2,\ldots,i_{n_c}(t)^2]^\top,
\end{align}
where $d(\phi,t)$ is an unmeasured torque disturbance that depends on time and position, i.e., \begin{equation}\label{eq:disturbances}
d(\phi,t) := d_1(t) + d_2(\phi).
\end{equation}
Neither the torque $T$ nor the coil inductances $L_c$ are measured. It is explained next how the torque $T$ of SRMs is controlled, given a model of $\hat{\mathbf{g}}$. 
\subsection{Control of Switched Reluctance Motors}
The nonlinear torque-current-angle relationship \eqref{eq:nonlin_plant} is inverted as follows. Given a model $\hat{\mathbf{g}}(\phi)$ of $\mathbf{g}(\phi)$, a commutation function $\mathbf{f}(\phi,T^*)$ is designed with \begin{equation}
\mathbf{f}(\phi,T^*) := \begin{cases}
\mathbf{f}^+(\phi) T^* & T^*\geq 0,\\
-\mathbf{f}^-(\phi) T^* & T^*<0,
\end{cases}
\end{equation}
where $T^*$ is the desired torque and $\mathbf{f}^+(\phi),\mathbf{f}^-(\phi): \mathbb{R}\to \mathbb{R}^{n_c}$. The functions $\mathbf{f}^-$ and $\mathbf{f}^+$ are designed to satisfy  \begin{align}\label{eq:requirements}
\hat{\mathbf{g}}(\phi)\mathbf{f}^+(\phi) &= 1, & \hat{\mathbf{g}}(\phi)\mathbf{f}^-(\phi) &= -1,\\
\mathbf{f}^+(\phi) &\geq \mathbf{0},&\mathbf{f}^-(\phi)&\geq \mathbf{0}.
\end{align}
Infinitely many functions $\mathbf{f}$ exist that satisfy these requirements, since $\mathbf{g}$ and $\mathbf{f}$ are vector functions. The control law $\mathbf{u}(t) = \mathbf{f}(\phi(t),T^*(t))$ then provides squared currents for every coil, given a certain desired torque $T^*$, see Figure \ref{fig:controlscheme}. Combining this control law with \eqref{eq:nonlin_plant} yields \begin{equation}\label{eq:truetorque}
T(\phi(t)) = \mathbf{g}(\phi(t)) \mathbf{f}(\phi(t),T^*(t))+d(\phi,t). 
\end{equation}
If the model is exact, with $\hat{\mathbf{g}}(\phi)=\mathbf{g}(\phi)$ and $\mathbf{f}$ meeting Requirement \eqref{eq:requirements}, then without disturbances, $T(\phi)=T^*(\phi)$ for all $\phi$. If not, the realized torque differs from the desired torque, leading to a problem known as torque ripple \cite{Gan2018}, impairing tracking performance. This underlines the importance of identifying $\hat{\mathbf{g}}(\phi)\approx \mathbf{g}(\phi)$, as detailed next.

\subsection{Problem formulation}
The aim is to develop a method for identifying the nonlinear torque-current-angle relationship $\hat{\mathbf{g}}(\phi)$ in SRMs, given measurements of the rotor position $\phi$ and squared currents $\mathbf{u}$, see Figure \ref{fig:controlscheme}. Key challenges include unmeasured produced torque $T$, unknown phase inductances $L_c$, and unknown linear dynamics $G(s)$ and disturbances $d(\phi,t)$.
\section{IDENTIFICATION OF SWITCHED RELUCTANCE MOTORS}\label{sec:method}
This section outlines the identification of SRM dynamics $\mathbf{g}(\phi)$, including the necessary assumptions, experimental design method, and posing of the estimation problem.
\subsection{System and model assumptions}
The nonlinear relationship ${\mathbf{g}}(\phi)$ involving rotor torque $T$, squared currents $\mathbf{u}$, and rotor angle $\phi$ is identified using $\mathbf{u}$ and $\phi$, with the challenge being the unmeasured torque $T$, see \eqref{eq:nonlin_plant}. The key idea is to obtain a data set in which the torque $T$ is known to be approximately constant, i.e., \begin{equation}\label{eq:Tconst}
T(t) \approx T_{\textnormal{const}}.
\end{equation}
To satisfy this requirement, we assume the following. 
\begin{assumption}\label{ass:damp}
The sub-system $G(s)$, see Figure \ref{fig:controlscheme}, is linear and time-invariant, and contains exactly one integrator. 
\end{assumption}
Assumption \ref{ass:damp} implies that constant rotor velocity leads to constant true torque $T(t)$. Consequently, if experiments can be designed in which the rotor velocity is constant, then the true torque acting on the rotor is equal to a certain $T_{\textnormal{const}}$. 

To achieve a constant rotor velocity for data collection, closed-loop experiments are performed using {a fixed} feedback controller $C(s)$ and a by-design imperfect commutation function $\mathbf{f}_{\text{imp}}$. Observe from \eqref{eq:truetorque} that the relationship \begin{equation}\label{eq:lsqr1}
\mathbf{g}(\phi(t)) \mathbf{f}_{\text{imp}}(\phi(t),T^*(t)) = T_{\text{const}}-d(\phi,t),
\end{equation}
has $\mathbf{g}$ and $d$ as unknowns. While $T_{\text{const}}$ is not exactly known because $G(s)$ is unknown, it is known to be constant. As explained later, its exact value is irrelevant. Expression \eqref{eq:lsqr1} can be interpreted as having a disturbed observation of $T_{\text{const}}$, following from the product of $\mathbf{g}(\phi)$, which is to be identified, and the user-defined $\mathbf{f}_{\text{imp}}(\phi,T^*)$. Moreover, the following assumptions are imposed.
\begin{assumption}\label{ass:period}
Disturbances $d(t,\phi)$ are not spatially periodic with period $p = \frac{2\pi}{k n_t},\ k\in\mathbb{N}$.
\end{assumption}
 \begin{assumption}\label{ass:sensornoise}
    The rotor position $\phi$ is measured exactly.  
\end{assumption}
Finally, $\hat{\mathbf{g}}$ is parametrized linearly in parameters $\boldsymbol{\theta}\in\mathbb{R}^{n_{{\theta}}}$ as \begin{equation}\label{eq:glinparam}
\hat{\mathbf{g}}^\top(\phi,\boldsymbol{\theta}) = \boldsymbol{\psi}_g(\phi) \boldsymbol{\theta},
\end{equation} 
with basis $\boldsymbol{\psi}_g:\mathbb{R}\to \mathbb{R}^{n_c\times n_{\boldsymbol{\theta}}}$. Example parametrizations are a Fourier basis or periodic radial basis functions \cite{Duvenaud2014}. In this paper, the focus is restricted to a Fourier basis of the form \begin{equation*}
\boldsymbol{\beta}(\phi)\hspace{-0.05cm} = \hspace{-0.05cm}[1, \sin(n_t\phi),\cos(n_t\phi),\ldots,\sin(n_h n_t\phi),\cos(n_h n_t\phi)],
\end{equation*}
 where $n_h$ is the number of harmonics, such that \begin{equation}
\boldsymbol{\psi}_g(\phi) = \mathbf{I}_{n_c} \otimes \boldsymbol{\beta}(\phi){,}
\end{equation} {where $\otimes$ denotes the Kronecker product.} This leads to a parametrization of $\hat{\mathbf{g}}$ with $n_{{\theta}} = n_c(1+2 n_h)$ parameters, in which all rotor teeth of the SRM are assumed to be identical. 

Next, experiment design in view of requirements \eqref{eq:requirements} and \eqref{eq:Tconst} is investigated.
\subsection{Experiment design}
This subsection covers obtaining constant rotor velocity samples through closed-loop experiments, explaining the choice of feedback controller $C$, and detailing commutation functions $\mathbf{f}_{\text{imp}}$. In contrast to the standard experiment design framework in closed-loop {\cite{gevers2011optimal}}, the goal here is to manipulate the commutation functions $\mathbf{f}_{\text{imp}}$ to generate sufficient excitation for learning the parameters of $\mathbf{g}$.
\subsubsection{Feedback control design}
Feedback control is applied to satisfy Requirement \eqref{eq:Tconst} during experiments.
When only an imperfect commutation function $\mathbf{f}_{\text{imp}}$ is available such that $T\neq T^*$, then a {ramp} reference $\phi_r$ with $\textnormal{d}\phi_r/\textnormal{d}t=\omega_r\in \mathbb{R}$ is still accurately tracked using a stabilizing feedback controller $C(s)$. Practical considerations of the choice of $C(s)$ and $\omega_r$ that leads to accurate tracking of a ramp reference are addressed in Section \ref{sec:implementation}. 
\subsubsection{Commutation function design}
$N_{\textnormal{exp}}$ experiments are designed, each with a different commutation function $\mathbf{f}_{\text{imp}}^{(i)}$, balancing the following requirements: \begin{enumerate}
    \item $\mathbf{f}_{\text{imp}}^{(i)}$ should invert $\mathbf{g}$ sufficiently well (see \eqref{eq:requirements}) so as not to destabilize the closed loop. 
    \item $\mathbf{f}_{\text{imp}}^{(i)}$ should not invert $\mathbf{g}$ perfectly, i.e., equality should not hold in \eqref{eq:requirements}, since then $\mathbf{g}(\phi) \mathbf{f}_{\text{imp}}^{(i)}(\phi,T^*)$ in \eqref{eq:lsqr1} is approximately constant for a constant feedback control effort $T^*(t)$. This leads to a problem of persistence of excitation, as is detailed in Section \ref{sec:persistence}. 
\end{enumerate}
Neither requirement can be verified for a given $\mathbf{f}_{\text{imp}}^{(i)}$ a priori because the true $\mathbf{g}$ is not available. Hence, an iterative approach is taken to design $\mathbf{f}_{\text{imp}}^{(i)}$. 

First, assume that some imperfect, simple model $\hat{\mathbf{g}}_{s}$ is available, i.e., a sinusoid per coil in the form of \begin{equation}\label{eq:g_simple}
\hat{{g}}_{s,c}(\phi,\phi_o) = \sin \left(N_t \phi + \frac{2\pi(c-1)}{n_c}+\phi_{o}^{(i)}\right).
\end{equation}
This model assumes a sinusoidal relationship between the torque-current ratio and the rotor angle. The coils are assumed to be equidistantly distributed along the rotor teeth and the offset $\phi_o^{(i)}$ is a parameter chosen differently for each experiment $i$, as detailed later. The commutation functions $\mathbf{f}_{\text{imp}}^{(i)}$ that invert $\hat{\mathbf{g}}_{s}$ are then chosen as \begin{align}\label{eq:fimp}
{f}_{\text{imp}, c}^{(i)}\left(\phi,T^*,\phi_o^{(i)}\right)=&{f}_{\mathrm{TSF},c}\left(\phi+\frac{2 \pi(c-1)}{n_c}+\phi_o^{(i)},T^*\right)\nonumber \\
&\cdot\operatorname{sat}\left(1 / \hat{{g}}_{s,c}\left(\phi,\phi_o^{(i)}\right)\right)T^*,\\
\textnormal{sat}(x) :=&\begin{cases}x_{\min}  & x < x_{\min}, \\
{x}  & {x_{\min}\leq x \leq x_{\max},} \\
x_{\max} & x > x_{\max},\end{cases}
\end{align}
where ${f}_{\text{imp}, c}^{(i)}$ refers to the $c^{\mathrm{th}}$ element of $\mathbf{f}_{\text{imp}}^{(i)}$. Moreover, $\mathbf{f}_{\mathrm{TSF}}(\phi,T^*): \mathbb{R}\times \mathbb{R}\to\mathbb{R}^{n_c}$ is a torque sharing function that divides a desired torque to different coils \cite{Wang2016}, satisfying \begin{equation}
\sum_{c=1}^{n_c} {f}_{\mathrm{TSF},c}(\phi,T^*) = \begin{cases}
    1 & T^* \geq 0,\\
    -1 & T^* < 0.
\end{cases}
\end{equation} Note that at values of $\phi$ where $\mathbf{g}_c(\phi)=0$, $\mathbf{f}_{\mathrm{TSF},c}(\phi,T^*)=0$ by design, and hence, \eqref{eq:fimp} is well defined for all $\phi$. 

The full data collection procedure is described in Procedure \ref{proc:datacol}. Experiments utilize various model offsets $\phi_o$, discarding any with error $|e|$ surpassing the threshold $e_{\text{safety}}$, which could indicate instability. If the closed-loop is stable but $|e|$ exceeds a user-defined maximum $e_{\text{max}} < e_{\text{safety}}$, then Requirement \eqref{eq:Tconst} is violated so the procedure is restarted for a slower reference velocity. 
Finally, the experiments are repeated in the other direction. 
\makeatletter
\renewcommand*{\ALG@name}{Procedure}
\makeatother
\afterpage{
\vspace*{-10pt}
\begin{algorithm}
\caption{Closed-loop data collection}\label{proc:datacol}
\begin{algorithmic}[1]
	\Require 
 Step size $\delta$, limit $\phi_{r,\text{N}}$.
	\State Define $\omega_r>0$, initialize $\phi_o^{(1)}\leftarrow \phi_{o,\min}$, $i\leftarrow 1$.
        \While {$\phi_o^{(i)} < \phi_{o,\max}$}
                    \State Perform experiment with $r=\phi_r$ and $\phi_o^{(i)}$, see \eqref{eq:fimp}. 
            \If{$|e| > e_{\text{max}}$}
                \State {Decrease {$|\omega_r|$}, return to Step 2}.
            \ElsIf{$|e| > e_{\text{safety}}$}
                \State Skip to Step 10. 
            \EndIf
            \State Store samples $\{T^*(t_j),\mathbf{u}(t_j),\phi(t_j)\}_{j=1}^N$.
            \State $\phi_o^{(i+1)}\leftarrow \phi_o^{(i)} + \delta$, $i\leftarrow i+1$.
        \EndWhile
        \State Set $\phi_r\leftarrow -\phi_r$ and repeat Steps 2-11.
	\end{algorithmic}
\end{algorithm}}
\subsection{Bayesian identification}\label{sec:sol}
A linear model is considered in the form \begin{equation}\label{eq:linearmodel}
\mathbf{b} = \mathbf{X} \boldsymbol{\theta} - \mathbf{d}.
\end{equation}
Here, $\mathbf{d}=\mathbf{d}_1+\mathbf{d}_2$ represents a vector that contains all (unmeasured) values of $d(\phi(t_k),t_k)$, see \eqref{eq:disturbances}, and $\mathbf{b}$ and $\mathbf{X}$ are defined below to represent a matrix-vector reformulation of \eqref{eq:lsqr1}, with $\hat{\mathbf{g}}=\hat{\mathbf{g}}(\phi,\boldsymbol{\theta})$. 

First, observe from \eqref{eq:lsqr1} that $\mathbf{b}$ should contain the samples $T_{\text{const}}$ with the appropriate sign. For ease of notation, but without loss of generality, we assume that an equal number of forward experiments $N_{\text{exp}}$ and backward experiments are successfully carried out, each with $N$ samples, such that $N_{\text{tot}}=2N_{\text{exp}}N$. This results in \begin{equation}\label{eq:b}
\mathbf{b} = T_{\text{const}} [{\mathbf{1}}_{N_{\text{exp} N}}^\top\  -{\mathbf{1}}_{N_{\text{exp} N}}^\top ]^\top.
\end{equation}
Note that the exact value of $T_{\text{const}}$ is irrelevant since $\hat{\mathbf{g}}$ is linear in $\boldsymbol{\theta}$: if it deviates by a constant factor, then the resulting $\hat{\mathbf{g}}$ {deviates} by the same factor. This is no problem since it is effectively a different loop gain of the linearized system, which can be compensated for by re-scaling $C(s)$. Hence, $T_{\text{const}}$ is chosen as \begin{equation}
T_{\text{const}} = \frac{1}{N_{\text{tot}}}\sum_{k=1}^{N_{\text{tot}}} |T^*(t_k)|.
\end{equation}
Next, the design matrix $\mathbf{X}$ in \eqref{eq:linearmodel} is constructed. Observe from \eqref{eq:lsqr1} that one element of $\mathbf{X}\boldsymbol{\theta}$ must describe a sample $\hat{\mathbf{g}}(\phi(t),\boldsymbol{\theta})\mathbf{f}_{\text{imp}}(\phi(t),T^*(t))$. To obtain $\mathbf{X}$, first define \begin{align}
\boldsymbol{\Psi}_g &= [\boldsymbol{\psi}_g^\top(\phi_1),\ldots,\boldsymbol{\psi}_g^\top(\phi_{N_{\text{tot}}})]^\top,\\ 
\mathbf{U} &= \sum_{i=1}^{N_{\text{tot}}} \mathbf{E}_{ii} \otimes \mathbf{u}_i^\top,
\end{align}
where $\mathbf{u}_i\in\mathbb{R}^{n_c}$ is the $i^{\mathrm{th}}$ vector of squared currents in the data set and $\mathbf{E}_{ii}$ is a matrix unit, i.e., a matrix with only one nonzero entry with value 1 at the $i^{\mathrm{th}}$ row and column. The design matrix $\mathbf{X}$ is then obtained as \begin{equation}\label{eq:X}
\mathbf{X} = \mathbf{U} \boldsymbol{\Psi}_g.
\end{equation}
Having defined the linear model, we now pose the estimation problem and apply a Bayesian framework to take into account the uncertain prior knowledge of $d(\phi,t)$. First, a Gaussian prior is posed on $\boldsymbol{\theta}$: \begin{equation}
\boldsymbol{\theta}\sim\mathcal{N}(\mathbf{0},\mathbf{I}).
\end{equation}
In addition to this, we pose Gaussian priors on the disturbances $d_1$ and $d_2$ as defined in \eqref{eq:disturbances}, namely, \begin{equation}
\begin{aligned}
\mathbf{d}_1 &\sim \mathcal{N}(\mathbf{0},\sigma^2 \mathbf{I}),\\
\mathbf{d}_2 &\sim \mathcal{N}(\mathbf{0},\boldsymbol{\Sigma}).
\end{aligned}
\end{equation}
These priors state that the temporal disturbances are assumed i.i.d. white noise, and the spatial disturbances are a priori expected to be described by some covariance matrix $\boldsymbol{\Sigma}$. The prior variance on the spatial disturbances is defined as \begin{equation}\label{eq:kernel}
\text{cov}(d_2(\phi),d_2(\phi')) = k(\phi,\phi'),
\end{equation}
where $k$ is a kernel function that can be chosen in such a way as to encode prior information of $d_2(\phi)$, e.g., periodicity with a known spatial frequency, see \cite{Duvenaud2014} for details. 

Recall from Assumption \ref{ass:sensornoise} that for all samples in the data set at time $t_k$, a noiseless measurement of $\phi(t_k)$ is available. Hence, the covariance matrix $\boldsymbol{\Sigma}$ can be constructed as \begin{equation}
\boldsymbol{\Sigma}_{ij} = k(\phi(t_i), \phi(t_j)),
\end{equation} 
where $i,j\in\{1,\ldots,N_{\text{tot}}\}$. 

With these priors, the estimate $\hat{\boldsymbol{\theta}}$ and model variance are given \cite[Section 4.2]{Pillonetto2022} by \begin{equation}\label{eq:bayes}
\begin{aligned}
\hat{\boldsymbol{\theta}}=\mathbb{E}[\boldsymbol{\theta}|\mathbf{X},\mathbf{b}]& =\mathbf{X}^\top (\mathbf{X}\mathbf{X}^\top + \boldsymbol{\Sigma}+\sigma^2 \mathbf{I})^{-1} \mathbf{b},\\
\text{Var}(\boldsymbol{\theta}|\mathbf{X},\mathbf{b}) &= \mathbf{I} - \mathbf{X}^\top(\mathbf{X} \mathbf{X}^\top + \boldsymbol{\Sigma}+\sigma^2 \mathbf{I})^{-1} \mathbf{X}.
\end{aligned}
\end{equation}
Finally, the model $\hat{\mathbf{g}}(\phi,\boldsymbol{\theta})$ is obtained from \eqref{eq:glinparam}. The next section addresses the implementation aspects of the developed identification method. 
\section{IMPLEMENTATION ASPECTS}\label{sec:implementation}
This section details the design of the feedback controller, persistence of excitation, and choice of reference.
W\subsection{Design of the linear feedback controller}
When $C(s)$ is designed with an integrator, a ramp reference can be accurately tracked. This is explained as follows.
First, ignoring the nonlinear feedback interconnection of $d$, the tracking error is given by \begin{equation}
{E}(s) = \frac{{R}(s)}{1+C(s)G(s)} = \frac{{\omega_r}}{s^2 (1+C(s)G(s))},
\end{equation}
{where we have considered the constant velocity reference $R(s)=\omega_r/s^2$.} For ease of analysis, it is assumed that the feedback controller is designed in continuous time before it is discretized for implementation. By Assumption \ref{ass:damp}, and the fact that $C(s)$ has one integrator, it follows that \begin{equation}
E(s) = \frac{\omega_r}{s^2 + L_{0}(s)},
\end{equation} 
where $L_{0}(s) := {s^2}C(s)G(s)$ has no integrators. From the final value theorem, it follows that the steady-state tracking error in the absence of disturbances is \begin{equation}\begin{aligned}
\lim_{t\rightarrow \infty} e(t) &= \lim_{s\rightarrow 0}s {E}(s) \\
&= \lim_{s\rightarrow 0} \frac{{\omega_r} s}{s^2 + L_{0}(s)} = 0,
\end{aligned}
\end{equation}
i.e., a ramp reference is tracked accurately when $C(s)$ contains an integrator.
\subsection{Persistence of excitation}\label{sec:persistence}
The offsets $\phi_o$ in Procedure \ref{proc:datacol} are the tuning parameters for the by-design imperfect commutation functions $\mathbf{f}_{\text{imp}}$ during experiment design, and these are related to persistence of excitation as follows.
 When $d(\phi,t)=0$ and this is incorporated as prior knowledge through $\sigma=0,\boldsymbol{\Sigma}=\mathbf{0}$, \eqref{eq:bayes} reduces to an ordinary least squares estimate. It follows that $\mathbf{X}$ must have rank $n_{\theta}$ for persistence of excitation. When insufficient values of $\phi_{o}$ are used, $\mathbf{X}$ can have linearly dependent rows through \eqref{eq:X}. Hence, a sufficient number of different experiments is required. 
\subsection{Reference design}
The design of the reference consists of $1)$ the chosen velocity $\omega_r$, and $2)$ the total stroke $\phi_{r,N}$. 
\subsubsection{Choosing the reference velocity}
The tracking error as a result of $d(\phi,t)$, again ignoring the nonlinear feedback interconnection, is given by the process sensitivity\begin{equation}
\frac{{E}(s)}{{D}(s)} = \frac{G(s)}{1+C(s)G(s)}.
\end{equation}
 As the frequency tends to zero, using the fact that $C(s)$ and $G(s)$ each have one integrator, we have
\begin{equation}\label{eq:process_limit}
\lim_{s\rightarrow 0} \frac{{E}(s)}{{D}(s)} = C^{-1}({0}) = 0.
\end{equation}
By choosing ${\omega_r}$ to be small, the position-dependent component of $d(\phi,t)$ will evolve slowly in time, and consequently, from \eqref{eq:process_limit} it follows that the tracking error as a result of these disturbances is small. 

A smaller ${\omega_r}$ requires more experimental time for the same stroke. When ${\omega_r}$ is too high, Requirement \eqref{eq:Tconst} ceases to hold. A heuristic that has proven effective in simulations is to choose $\omega_r$ such that $\|e\|_\infty \ll \frac{2\pi}{n_t}$, e.g., $\|e\|_\infty < 10^{-4}\frac{2\pi}{n_t}$.
\subsubsection{Choosing the reference stroke}
When the magnitude of $d_2(\phi)$ is large or it varies slowly with position, it is recommended to choose a larger stroke $\phi_{r,N}$. Indeed, by Assumption \ref{ass:period} and the fact that $\hat{\mathbf{g}}$ is spatially periodic with period $\frac{2\pi}{n_t}$, disturbances $d_2(\phi)$ leak to all relative angles mod$(\phi,\frac{2\pi}{n_t})$. This facilitates the choice of an accurate prior $k$ in \eqref{eq:kernel}, as illustrated in Section \ref{sec:simresults}. Moreover, when the rotor teeth vary significantly, it is also recommended to choose a larger stroke, since it is desired to obtain a $\hat{\mathbf{g}}$ that describes the average tooth. 
\section{SIMULATION RESULTS}\label{sec:simulation}
In this section, a simulation example is presented. First, the setting is explained, and subsequently, the results are shown. 
\subsection{Setting}
An SRM model with \(n_t=131\), \(n_c=3\), and linear dynamics \(G(s) = {1}/{(s^2+s)}\) is considered, sampled at \(1\) kHz. The true $\mathbf{g}$ consists of a sum of five sines and cosines per coil. A PID controller is given with a 20 Hz bandwidth and a position-dependent mechanical disturbance is present: \begin{equation}
d(\phi,t) = d_1(t) + 5\cdot 10^{-4}\sin\left(\frac{n_t}{1.4}\phi\right),
\end{equation}
with $d_1(t)\sim\mathcal{N}(0,7\cdot 10^{-9})$. The spatial frequency $\frac{n_t}{1.4}$ is unknown. The initial model $\hat{\mathbf{g}}_{s}$ is defined as in \eqref{eq:g_simple}. Procedure \ref{proc:datacol} is followed for $\phi_{o}\in\{-0.2,0.2\}$ and a reference velocity of $\omega_r=10$ mrad/s. The total reference stroke in each experiment is approximately 12 teeth. The data during the first two teeth are removed to exclude the transient and the rest of the data is downsampled to leave a total of $N=1000$ samples per experiment. 

\subsection{Results}\label{sec:simresults}
\begin{figure}[tb]
    \centering
        \vspace{3pt}
    \input{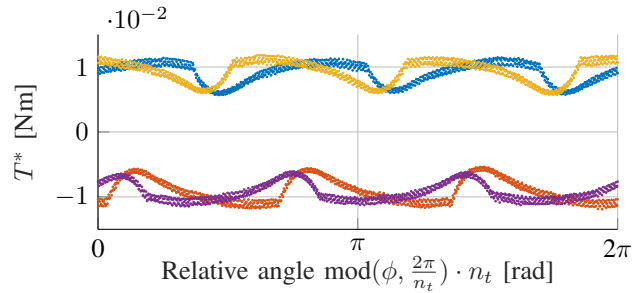}
    \caption{Desired torque samples $T^*$ plotted along one tooth for various simulations, each using a different $\mathbf{f}_{\text{imp}}$. The feedback controller adjusts $T^*$ to minimize tracking error for each $\mathbf{f}_{\text{imp}}$, influenced by $d(\phi,t)$.}\label{fig:tstar_sim}
\end{figure}

During the experiments, the maximum error is $\|e\|_{\infty}\approx 5\cdot 10^{-7}$ rad. Since the spatial period of a tooth is $\frac{2\pi}{131}\approx 5\cdot 10^{-2}$ which is five orders of magnitude larger than the error, this suggests that Requirement \eqref{eq:Tconst} is satisfied. Figure \ref{fig:tstar_sim} shows the desired torque during experiments, with each commutation function \(\mathbf{f}_{\text{imp}}\) introducing torque ripple, mostly compensated by \(T^*(t)\) computed by \(C(s)\).

If known, the spatial frequency of the disturbances can be included in the prior through \(k(\phi,\phi')\), see \eqref{eq:kernel}. Instead, realize that $\hat{\mathbf{g}}$ is spatially periodic with the period of a tooth, but the disturbances are not (see Assumption \ref{ass:period}). This means that when sufficient teeth are observed, $d_2(\text{mod}(\phi,2\pi/n_t))$ can be approximated as i.i.d. white noise. This is also visible from Figure \ref{fig:tstar_sim}, where $T^*(t)$ is compensating for $d(t)$. Hence, the prior on $d_2(\phi)$ is simply chosen as i.i.d. white noise using $k(\phi,\phi') = 10^{-6}\delta_{\phi\phi'}$ with \begin{equation}
\delta_{\phi\phi'}:=\begin{cases}
    1 & \phi = \phi',\\
    0 & \text{otherwise}.
\end{cases}
\end{equation} This leads to a diagonal $\boldsymbol{\Sigma}$, so $d_2$ is lumped together with $d_1$, and we choose $\sigma=0$. 

\begin{figure}[tb]
    \centering
\input{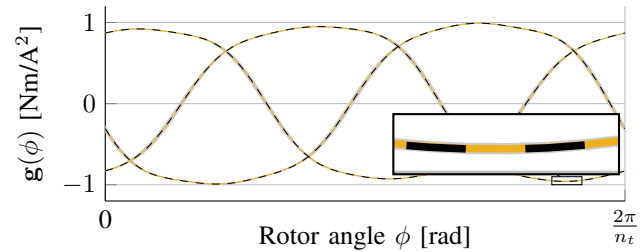}
\caption{True $\mathbf{g}$ (\protect\blackdash) in simulation, estimate $\hat{\mathbf{g}}$ (\protect\yelline) and 95\% uncertainty bounds (\protect\greyarea). An accurate model is obtained with a small variance.}
    \label{fig:g_sim}
\end{figure}

The resulting estimate $\hat{\mathbf{g}}$ from \eqref{eq:bayes} is depicted in Figure \ref{fig:g_sim}. The model \(\hat{\mathbf{g}}\) closely resembles the true \(\mathbf{g}\), even with significant position-dependent disturbances \(d\), in just four one-minute experiments.
\section{EXPERIMENTAL RESULTS}\label{sec:experiments}
In this section, experimental results are presented. The setting is described first, and the results are shown afterward. Finally, the validation of the results is addressed. 
\subsection{Setting}
A real SRM from TNO \cite{Kramer2020} is considered with $n_t=131$ and $n_c=3$. The frequency response of $G(s)$ is measured using an imperfect commutation function, leading to a linear model sufficient for the design of a PID controller with 20 Hz bandwidth. The models $\hat{\mathbf{g}}_s$ for Procedure \ref{proc:datacol} are given by \eqref{eq:g_simple} using $\phi_o\in\{-0.7,-0.65,-0.23, 0\}$. Six experiments were conducted at \(\omega_r=7.5\cdot 10^{-4}\) rad/s, each with a total stroke of roughly 10 teeth. It is emphasized that in this experimental setting, the disturbances $d(\phi,t)$ are unknown and the true function $\mathbf{g}(\phi)$ is not exactly periodic, since the rotor teeth slightly differ due to manufacturing tolerances. The transient is removed from the data, and the dataset is downsampled to obtain a total of $N=1157$ samples per experiment. 
\subsection{Results}
\begin{figure}[tb]
\centering
    \vspace{3pt}
\input{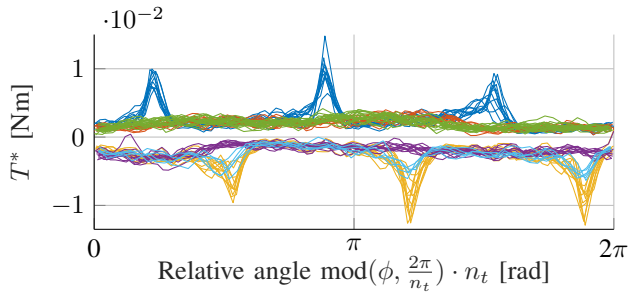}
\caption{Experimental data. The desired torque $T^*(t)$ is plotted for one tooth, with colors representing different experiments and commutation functions $\mathbf{f}_{\text{imp}}$. The pattern illustrates how $T^*(t)$ compensates for imperfections in each $\mathbf{f}_{\text{imp}}$, with noticeable tooth-by-tooth variations.}\label{fig:exp_Tstar}
\end{figure}

With a peak error of \(\|e\|_\infty\approx 9\) $\mu$rad during data acquisition, four orders of magnitude smaller than the rotor teeth, Requirement \eqref{eq:Tconst} appears satisfied.
Figure \ref{fig:exp_Tstar} depicts the resulting $T^*$, showing that a large $T^*$ is required to offset the underestimated current generated by $\mathbf{f}_{\text{imp}}$. Despite the model assuming identical teeth, there are noticeable variations between individual teeth. However, since the objective is to design a commutation function for the average tooth, the model structure is retained, and the increased variance in the estimate is accepted.

\begin{figure}[tb]
\centering
\input{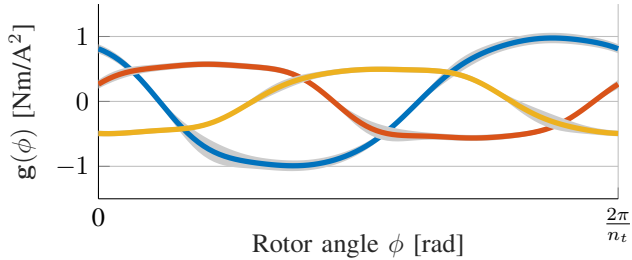}
\caption{Resulting model $\hat{\mathbf{g}}$ from the experimental data for each coil (\protect\blueline,\protect\redline,\protect\yelline) with 95\% uncertainty bounds (\protect\greyarea). As expected, the three functions are shifted approximately 120$^\circ$ in phase. The blue coil produces more torque than the other coils, possibly because of a smaller air gap.}\label{fig:g_exp}
\end{figure}

Similarly to Section \ref{sec:simulation}, no prior information on the disturbances is used, and the choice $k(\phi,\phi')=3\cdot 10^{-6}\delta_{\phi\phi'}$ is made. The resulting model is shown in Figure \ref{fig:g_exp}. There are no torque sensors in the experimental setup, which means that validation is not trivial, as is explained in the next section. 
\subsection{Validation}
The model \(\hat{\mathbf{g}}\) appears accurate for two reasons. Firstly, the first blue coil in Figure \ref{fig:g_exp} provides more torque than the others.
 This is consistent with the observation that the feedback controller $C(s)$ decreases the desired torque $T^*$ whenever a current is sent to this coil, to achieve low tracking error. This suggests that this coil truly does produce more torque than the others, possibly because the air gap is smaller. 

Second, when $\hat{\mathbf{g}}$ is used to design an inverting commutation function $\mathbf{f}$, the error is reduced significantly with respect to an $\mathbf{f}$ designed using a $\hat{\mathbf{g}}$ in which only the first harmonic is non-zero. For tasks with $\omega_r<0.5$, $\|e\|_2$ is reduced by an order of magnitude. Note that unless $\hat{\mathbf{g}}=\mathbf{g}$ holds exactly, the tracking error is highly dependent on the specific shape of $\mathbf{f}$, even though $\hat{\mathbf{g}}\mathbf{f}=1$ holds. Therefore, the identification method should be applied to an SRM equipped with torque sensors, to better quantify the achieved performance increase. 

\section{CONCLUSIONS}\label{sec:conclusions}
An identification method for Switched Reluctance Motors is developed that accurately captures the relationship between rotor torque, rotor angle, and currents. The approach does not rely on torque measurements or complex models and is therefore easily deployed in a production context, without the need for a dedicated calibration setup. A simulation example has shown that the method is robust to position-dependent disturbances, and experimental results from a real SRM confirm that the identification method enables commutation designs that lead to significantly better tracking performance. 

Further research will be aimed at further experimental validation. Moreover, the application of linear learning control techniques for experiment design will be explored, to overcome the need for slow movements during data collection. 

\addtolength{\textheight}{-12cm}   




\bibliography{sysid_g/library}

\end{document}